\documentclass[11pt]{article}
\usepackage{authblk}
\usepackage[utf8]{inputenc}
\usepackage[T1]{fontenc}
\usepackage{graphicx}
\usepackage{amsmath}
\usepackage{amssymb}
\usepackage{hyperref}
\usepackage[margin=25mm]{geometry}
\usepackage{parskip}

\usepackage{etoolbox}
\usepackage[table]{xcolor}
\usepackage{pgf}
\usepackage{colortbl}
\usepackage{multirow}
\usepackage{diagbox}
\usepackage{longtable}

\providecommand{\doi}[1]{\href{https://doi.org/#1}{doi:#1}}

\definecolor{low}{HTML}{76f013}
\definecolor{middle}{HTML}{fce303}
\definecolor{high}{HTML}{ec462e}

\author[1,2]{A.~V.~Smirnov}
\author[1]{V.~A.~Fokin}
\author[1]{E.~Iu.~Chuvashov}
\affil[1]{Lomonosov Moscow State University}
\affil[2]{Moscow Center for Fundamental and Applied Mathematics}
\affil[ ]{\{asmirnov@srcc.msu.ru, v.fokin2004@gmail.com, wowonline.fit@mail.ru\}}
\date{\today}
\title{Factorization of denominators as a `fuel' for Feynman integral reduction}
\begin{document}
\maketitle

\begin{abstract}
Rational-function simplification is key bottlenecks in integration-by-parts (IBP) reduction of Feynman integrals. We study denominator factorization
          patterns appearing in IBP coefficients and develop practical algorithms for extracting and exploiting factorized denominator structure within the {\tt FUEL} interface. The resulting workflow reduces reconstruction cost and improves robustness of large-scale reductions.
\end{abstract}


\section{Introduction}
\label{sec:intro}

The evaluation of Feynman integrals is a central task in perturbative quantum field theory calculations. It is based on the integration-by-parts reduction approach \cite{Chetyrkin:1981qh, Tkachov:1981wb}, which makes it possible to represent all integrals belonging to a given family (defined by a particular set of propagators) in terms of a finite set of simpler integrals known as master integrals.

Normally, a reduction means combining IBP identities in nontrivial ways. A systematic procedure for this is the so-called Laporta algorithm \cite{Laporta:2000dsw}. Several
public implementations of this algorithm exist, including {\tt AIR}
\cite{AIR}, {\tt Reduze} \cite{Studerus:2009ye, Reduze}, {\tt LiteRed}
\cite{Lee:2012cn, LiteRed2}, {\tt FIRE}
\cite{Smirnov:2008iw, Smirnov:2014hma, Smirnov:2024onl, Smirnov:2023yhb, Smirnov:2025prc}, and {\tt Kira} \cite{lange2025kira3integralreduction, Laporta:2000dsw, Lee:2012cn, LiteRed2, Lee:2013mka, Maierhofer:2017gsa}.

There are two competitive approaches to Feynman integral reduction, a classical one when one directly solves the system of equations and an alternative one with the modular approach when reduction is performed multiple times for fixed values of kinematic invariants and then a reconstruction of an analytic function is performed \cite{FIRE6, Klappert:2020nbg, Kira, Kleine2005, lange2025kira3integralreduction, Laporta:2000dsw, Lee:2012cn, LiteRed2, Lee:2013mka, Maierhofer:2017gsa, Reduze, vonManteuffel:2014ixa, Mokrov:2023vva, Monagan2004, Peraro:2016wsq, Peraro:2019svx, Symbolica, FIRE6, Smirnov:2013dia, Smirnov:2020quc, Smirnov:2008iw, Smirnov:2014hma, Smirnov:2024onl}.

In this paper, however, we will focus mainly on the classical approach where
the reduction procedure is technically solving a large sparse linear system
with polynomial coefficients. Still our method is only applicable within the
modular approach to speed up the reconstruction stage.

One can pay a close attention to structures appearing
in denominators of expressions of arbitrary Feynman integrals to master integrals.
The denominators can be considered as poles which, for Feynman integrals, can exist only for specific values
of kinematic invariants.
In particular, it is known (while not strictly proven) that one can always choose
such a basis of master integrals that the denominators can be factorized (or split)
into a product of a part depending on the spacetime dimension \(d\) and a part depending on other kinematic variables~\cite{Smirnov:2020quc, Usovitsch:2020jrk}.

Even more, the denominators tend to be repeating and arising in powers which leads
to an idea that it might be useful to keep the denominators factorized in order
to speed up the reduction. A common objection is that the factorization is an
algorithmically complicated operation and the need to perform it might be a performance degrade. But as was initially suggested by Ruijl, the author of {\tt Symbolica} \cite{Symbolica}, the need to factorize might happen not that often, so that there is a chance
to obtain performance gain. The {\tt Symbolica} library already has such an option for years that can be turned on by the {\tt FUEL} interface. However this fact seems to be
not widely known or used due to the fact that {\tt Symbolica} is a commercial package.

The idea of this paper is to implement denominator factorization with the use of
the {\tt FLINT} library \cite{flint}. The main problem that lies here is that
although {\tt FLINT} has an object representing a factorized polynomial
it has no arithmetical operations on such objects,
so one needs to implement a logic of working with such objects in FUEL \cite{Mokrov:2023vva}.
Moreover the communication between {\tt FUEL} and {\tt FIRE} consists of sending strings to each other, and such a string should represent a polynomial. In case one needs to keep
the factorized structure {\tt FIRE} should be able to store such structure after obtaining
it from {\tt FUEL} and then send it back, and the {\tt FUEL} parser should be able to interpret
the expression and not join the denominator factors. Thus implementing the idea
leads to a rather complicated code.

The goal of this paper is to describe how we implemented the factorization of
denominators with {\tt FLINT} in the {\tt FUEL} interface and how this can be used in {\tt FIRE}.
This also leads to a need to describe how we proposed and implemented a decomposed format for storing {\tt FLINT} expressions. We also provide benchmarks confirming the efficiency of such an approach for
problems with multiple kinematic invariants.

In the following section we are going to describe how we implemented the approach with factorization with the {\tt FUEL} interface with the use of {\tt FLINT}. In section~2 we provide instructions how to use the new approach with {\tt FIRE} and in section~3 we provide performance benchmarks and measurements.

\section{Implementation in {\tt FUEL}}
\label{sec:fuel}
To understand the following one needs to remind how {\tt FIRE} works with the {\tt FUEL} interface
and how other IBP reduction programs might use {\tt FUEL}. The main {\tt FUEL} evaluation function
receives a string containing an expression as an input and returns also a string with a simplified expression. Depending on options, this expression might have a human-readable setup (needed anyway before saving the final result) or any other form.
The only strict restriction that {\tt FIRE} imposes is that a pure zero result should always be returned as ``0''. {\tt FIRE} checks for an expression being equal to ``0'' and if it is not identically equal considers it
a non-zero rational function that can appear in a denominator. Now {\tt FIRE} works with these expressions, adding them to each other, multiplying, dividing, subtracting. But {\tt FIRE} does not have any built-in algebraic system, so the operations are pure string expressions. For example, after receiving expressions {\tt expr1} and {\tt expr2} from {\tt FUEL}, {\tt FIRE} can ``add them to each other'', resulting in {\tt (expr1) + (expr2)}. Afterwards such a combined expression can be sent back to the {\tt FUEL} parser, that is going to interpret it properly and to simplify. Thus {\tt FUEL} is free to use any optimized output for storing expressions unless {\tt FIRE} directly requests to return it in a human-readable form with a single restriction to return a pure zero as ``0''.

\subsection{Decomposed output without factorization}

To support the storage of factorized expressions and to speed up parsing, we introduced a {\tt decomposed output}
for polynomials. This format was first implemented in an unpublished coursework project~\cite{Chuvashov2025Serialization}.

In the standard {\tt FIRE}--{\tt FUEL} workflow rational functions are represented as fractions of multivariate polynomials
\[
  f(x_1,\dots,x_n) \;=\; \frac{p(x_1,\dots,x_n)}{q(x_1,\dots,x_n)} \,,
\]
where the polynomials are stored internally as sums of monomials. To make parsing and serialization
efficient, {\tt FUEL} does not print full symbolic expressions such as \((x + y^2 - 5 z^{10})/(2x + z^2)\),
but instead uses a \textit{decomposed output} format that exposes the monomial structure explicitly.

A monomial in \(n\) variables
\[
  m(x_1,\dots,x_n) \;=\; c\,x_1^{k_1} x_2^{k_2} \cdots x_n^{k_n},
  \qquad c \in \mathbb{Z},\; k_i \in \mathbb{N}_0,
\]
is serialized as a compact record containing the coefficient followed by all exponents,
\[
  m \;\longmapsto\; \langle c \,;\, k_1, k_2, \ldots, k_n \rangle,
\]
where in the actual output the entries are separated by dots.
The ordering and number of variables \((x_1,\dots,x_n)\) are fixed by the {\tt FIRE} problem definition,
so the variable names themselves do not have to be printed; zero exponents \(k_i = 0\) are kept
explicitly in the record so that the length of the exponent list is constant.

A polynomial
\[
  p(x_1,\dots,x_n) \;=\; \sum_{j} m_j(x_1,\dots,x_n)
\]
is then represented as a concatenation of monomial records,
\[
  p \;\longmapsto\;
  \langle c_1 \,;\, k_{1,1}, \ldots, k_{1,n} \rangle
  \;\langle c_2 \,;\, k_{2,1}, \ldots, k_{2,n} \rangle
  \;\cdots,
\]
with the same dot-separated encoding for each individual monomial as above.
A rational function \(p/q\) is printed as a pair of such sequences, enclosed in an additional
layer of angle brackets that separates the numerator and denominator. For example, the rational function
\[
  \frac{p(x,y,z)}{q(x,y,z)} \;=\;
  \frac{x + y^2 - 5 z^{10}}{2x + z^2}
\]
has the schematic decomposed output
\[
  \Big\langle
    \big\langle \langle 1.1.0.0 \rangle \,\langle 1.0.2.0 \rangle \,\langle -5.0.0.10 \rangle \big\rangle
    \;.\;
    \big\langle \langle 2.1.0.0 \rangle \,\langle 1.0.0.2 \rangle \big\rangle
  \Big\rangle.
\]
The outermost pair of brackets delimits the fraction, while the two inner bracketed blocks contain the
monomials of the numerator and denominator, respectively. In this representation {\tt FUEL} can build and
modify polynomials and rational functions incrementally, adding monomials “on the fly” without invoking
the slower general-purpose parser for human-readable algebraic expressions.

\subsection{Decomposed output with factorized denominators}

For factorized denominators, we extend the decomposed output format so that the internal
{\tt FLINT} representation \(\texttt{fmpz\_mpoly\_factor\_struct}\) can be reconstructed directly from
a single pass over the serialized string. Internally, a factorized polynomial
\[
  D(x_1,\dots,x_n)
  \;=\;
  c \prod_{i=1}^{r} p_i(x_1,\dots,x_n)^{e_i},
  \qquad c \in \mathbb{Q},\; e_i \in \mathbb{N}_0,
\]
is stored as an array of simple polynomial factors \(p_i\) together with their exponents \(e_i\)
and an overall rational constant \(c\).

In the factorized decomposed output, the numerator is still printed as a sequence of
monomials in the basic format described above. The denominator, however, is encoded in
a nested form that contains two flags, a rational constant, and the list of factors:
\[
  \Bigl\langle
    \text{flag}_1\,\text{flag}_2
    \Bigl\langle
      c_{\text{num}}.c_{\text{den}}
      \Bigl\langle
        e_1 \langle p_1^{\text{dec}} \rangle
        \cdots
        e_r \langle p_r^{\text{dec}} \rangle
      \Bigr\rangle
    \Bigr\rangle
  \Bigr\rangle
\]
where \(p_i^{\text{dec}}\) denotes the polynomial \(p_i\) written in decomposed form, and
\begin{itemize}
  \item \(\text{flag}_1\) and \(\text{flag}_2\) are two Boolean flags indicating whether
        factorization should be applied and whether the denominator is already stored
        in factorized form;
  \item \(c_{\text{num}}.c_{\text{den}}\) is the rational constant in front of the
        product of factors written as a numerator--denominator pair;
  \item each factor is written in the form \(e_i \langle p_i^{\text{dec}} \rangle\),
        where \(e_i\) is the exponent and \(p_i^{\text{dec}}\) is the corresponding
        polynomial factor in decomposed form:
        \[
          p_i(x_1,\dots,x_n)
          \;\longmapsto\;
          \langle c_{i,1} \,;\, k_{i,1,1}, \ldots, k_{i,1,n} \rangle
          \;\langle c_{i,2} \,;\, k_{i,2,1}, \ldots, k_{i,2,n} \rangle
          \;\cdots,
        \]
        with the same dot-separated encoding for each individual monomial as above.
\end{itemize}

Schematically, a rational function with a factorized denominator is serialized as
\[
  \Big\langle
    \underbrace{\langle \text{monomials of }N \rangle}_{\text{numerator}}
    \;.\;
    \underbrace{\langle
      \text{flag}_1\,\text{flag}_2
      \langle
        c_{\text{num}}.c_{\text{den}}
        \langle
          e_1 \langle p_1^{\text{dec}} \rangle
          \cdots
          e_r \langle p_r^{\text{dec}} \rangle
        \rangle
      \rangle
    \rangle}_{\text{factorized denominator}}
  \Big\rangle.
\]
During parsing, {\tt FUEL} reads the numerator monomials as before, then interprets the two flags
and the constant pair, and finally reconstructs the factors \(p_i\) together with their exponents
\(e_i\) to populate an \(\texttt{fmpz\_mpoly\_factor\_struct}\). This allows arithmetic
operations to exploit the factorized structure directly (e.g.\ cancelling common factors or updating exponents)
without repeatedly multiplying out large denominator polynomials.

For example, consider
\[
  \frac{N(x,y,z)}{D(x,y,z)}
  \;=\;
  \frac{x + y^2 - 5 z^{10}}
       {\frac{7}{5}\,(x-y)^2(2x+z^2)}.
\]
The numerator is printed in the same way as in the non-factorized case:
\[
  N(x,y,z)
  \;\longmapsto\;
  \langle 1.1.0.0 \rangle\,
  \langle 1.0.2.0 \rangle\,
  \langle -5.0.0.10 \rangle.
\]
For the denominator, the constant prefactor is written as
\[
  \frac{7}{5}
  \;\longmapsto\;
  7.5,
\]
and the two polynomial factors are written as
\[
  x-y
  \;\longmapsto\;
  \langle 1.1.0.0 \rangle\,
  \langle -1.0.1.0 \rangle,
  \qquad
  2x+z^2
  \;\longmapsto\;
  \langle 2.1.0.0 \rangle\,
  \langle 1.0.0.2 \rangle.
\]
So, with the actual syntax used in the code, the full serialized expression is
\[
  \Big\langle
    \big\langle
      \langle 1.1.0.0 \rangle\,
      \langle 1.0.2.0 \rangle\,
      \langle -5.0.0.10 \rangle
    \big\rangle
    \;.\;
    \big\langle
      11
      \big\langle
        7.5
        \big\langle
          2\big\langle
            \langle 1.1.0.0 \rangle\,
            \langle -1.0.1.0 \rangle
          \big\rangle
          1\big\langle
            \langle 2.1.0.0 \rangle\,
            \langle 1.0.0.2 \rangle
          \big\rangle
        \big\rangle
      \big\rangle
    \big\rangle
  \Big\rangle.
\]
Here the two digits \texttt{11} are two Boolean flags written one after another:
the first one is the value of \newline
 \texttt{factorize\_denominators}, and the second one
indicates whether the denominator is already stored in a factorized form.
The part \(\langle 7.5 \langle \cdots \rangle \rangle\) means that the overall
constant is \(7/5\), and after that the list of factors follows.

In particular, the block
\[
  2\big\langle
    \langle 1.1.0.0 \rangle\,
    \langle -1.0.1.0 \rangle
  \big\rangle
\]
should be read in the following way: the first number \(2\) is the exponent of the
factor, and the part inside brackets is the decomposed form of the polynomial
\[
  \langle 1.1.0.0 \rangle\,
  \langle -1.0.1.0 \rangle
  \;\longmapsto\;
  x-y.
\]
So this whole block represents the factor \((x-y)^2\).
In the same way,
\[
  1\big\langle
    \langle 2.1.0.0 \rangle\,
    \langle 1.0.0.2 \rangle
  \big\rangle
\]
represents the factor \((2x+z^2)^1\).
So each factor in the denominator is written as
\(e_i\langle p_i^{\text{dec}} \rangle\): first the exponent \(e_i\), then the
polynomial factor itself in the usual decomposed monomial format.

\subsection{Algorithms working with factorized denominators}

Since {\tt FLINT} does not support operations on expressions with factorized denominators, we had to implement out own logic of working with such functions. Now let us suppose that we have two expressions in a decomposed form with factorized denominators and that we need to perform an algebraic operation on them. To be more exact, let us take expressions

\begin{equation}
\frac{n}{d} = \frac{n}
{c \cdot d_1^{k_1} \cdot d_2^{k_2} \cdot \cdots \cdot d_m^{k_m}} \mbox{ and } \frac{n^\prime}{d^\prime} =  \frac{n^\prime}
{c^\prime \cdot d^{\prime k^\prime_1}_1 \cdot d^{\prime k^\prime_2}_2 \cdots \cdot d^{\prime k^\prime_l}_l}
\end{equation}

The following operations had to be considered.

\subsubsection{Multiplication}

We need to keep the expression in a proper format so that the product remains factorized and the numerator has no common factors with it. Also we need to avoid slow calls of factorization. First of all, we try to reduce the numerator $n$ with the factors of the denominator $d^\prime$. We try to reduce by each factor $d^\prime_i$ until its exponent allows, but only if this factor $d^\prime_i$ is not present among the factors of its own denominator $d$. This reduces the number of divisions, which is a very expensive operation. This is possible due to the fact that the input coefficients are already reduced, so the numerator $n$ cannot be divided by the factors of its own denominator $d$. Similarly, we reduce the numerator $n^\prime$ by factors of the denominator $d$. After the reduction, we multiply the remaining parts of the numerators forming the numerator of the result. The denominator of the result includes all the remaining factors of both denominators after the reduction. Here we take advantage of factorization by simply adding factors to the result. If a factor appears in both denominators, we simply add their exponents. The constants $c$ and $c'$ are multiplied, and the resulting constant $c'' = c''_{\text{num}} / c''_{\text{den}} = c \cdot c'$ is reduced so that $\gcd(c''_{\text{num}}, c''_{\text{den}}) = 1$. After that, we find the greatest common numerical constant $k$ of all monomials of the resulting numerator, compute $\gcd(k, c''_{\text{num}})$ and simplify the fraction. This prevents the growth of numerical coefficients and denominator constants during subsequent computations.

\subsubsection{Addition and subtraction}

Addition and subtraction are common operations differing only by the sign of the second expression. The basic formula for addition is $(n \cdot d^\prime + n^\prime \cdot d) / (d \cdot d^\prime)$. First, we bring fractions to a common denominator by multiplying the numerators of both fractions by the missing factors from the opposite denominators. After that, we add or subtract the numerators, forming an intermediate result. Similarly, we find the largest common numerical constant $k$ of all monomials of the resulting numerator, compute $\gcd(k, c''_{\text{num}})$ of $k$ and the constant of the resulting denominator $c''_{\text{num}}$, and simplify the fraction to keep numerical coefficients small. Next, we try to reduce the numerator by those factors of the resulting denominator that appeared in both original denominators with the same exponents. It makes no sense to try dividing by other factors because the input coefficients are already reduced: for example, consider $
\frac{n}{f \cdot d} + \frac{n^\prime}{d} = \frac{n + f \cdot n^\prime}{f \cdot d}$, Here $n$ is not divisible by $f$, while $f \cdot n^\prime$ is divisible by $f$, which means the sum will not be divisible by $f$.

\subsubsection{Division}

Division is the only operation in which you cannot do without calling factorization. Thankfully it is needed during reduction much less often that other operations. We factorise the numerator $n'$ and then work using the identity
$\frac{n}{d} / \frac{n^\prime}{d^\prime} = \frac{n}{d} \cdot \frac{d^\prime}{n^\prime}$, similar to the case of factorized multiplication. When reducing $d^\prime$ against the factors of $d$, we take advantage of the factorized form of $d^\prime$. We do not call expensive polynomial divisions. We simply reduce the exponents of common factors.

\subsubsection{Exponentiation}

Raising a coefficient to an integer power $e$ is relatively rare. If $e > 0$, we raise the numerator and the denominator constant to the power $e$, and multiply the exponents of the denominator factors by $e$. If $e$ is negative, we use $(\frac{n}{d})^e = (\frac{d}{n})^{-e}$. In this case we need to factorise the numerator $n$ and expand the denominator $d$ (i.e., convert it to a dense polynomial). Otherwise, it is similar to the case of a positive power.

\section{Usage in {\tt FIRE}}
\label{sec:fire}

In this small section we are going to provide instructions on how to use the described improvements within the {\tt FIRE} package. We assume that one already has {\tt FIRE7} installed. To get the possibility to use factorized denominators with {\tt FUEL} one has to switch to a release starting from 7.1. Then one runs {\tt make fuel} to get the latest submodule update, {\tt make clean} and {\tt make} to rebuild the code. Now {\tt FUEL} is ready to accept the new options.

The syntax is to use the {\tt calc\_options} option of {\tt FIRE} with \newline
{\tt -{}-calc\_options decomposed\_output,factorize\_denominators}.

One also can install {\tt Symbolica}, obtain a proper license and put it into the
{\tt ~/.symbolica/main\_license} file to be able to use it with {\tt --calc symbolica}. {\tt Symbolica} evaluator also accepts the {\tt factorize\_denominators},
{\tt decomposed\_output} is not required here and will have no effect.

\section{Benchmarks}
\label{sec:benchmarks}

\subsection{Benchmarks on physical examples}

In this section we are going to provide a number of benchmarks demonstrating the performance of the new approach. We are going to take a number of physical examples and compare the code in the following modes:

\begin{itemize}
 \item Default with {\tt flint};
 \item Decomposed output mode with {\tt flint};
 \item Factorized mode (together with decomposed output) {\tt flint}.
 \item With the use of {\tt symbolica} as a simplification library;
 \item Factorized mode with {\tt symbolica}.
\end{itemize}

The results below do not show a clear winner in the competition of out approach with {\tt FLINT} and {\tt Symbolica}, but clealry demonstate the efficiency of our approach compared with pure {\tt FLINT} runs.

\subsubsection{Two-loop massive nonplanar double box}
\label{subsec:npdbox}

\begin{figure}
  \centering
  \includegraphics[width=0.4\textwidth]{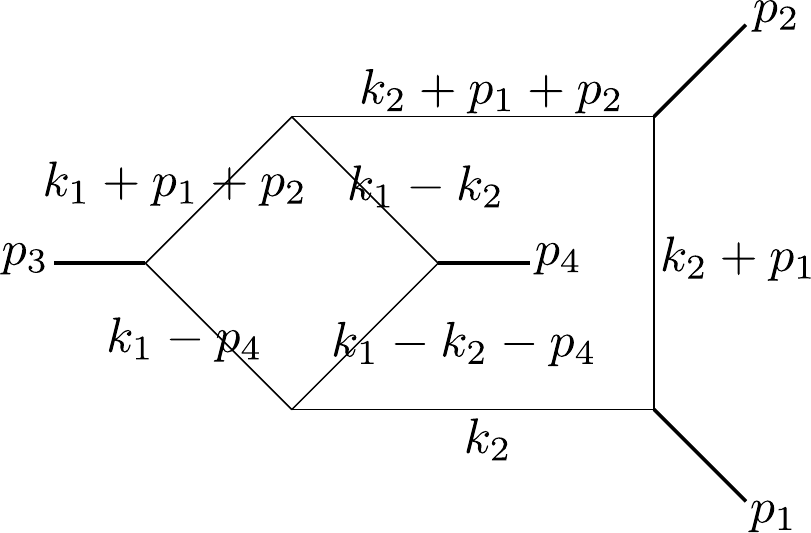}
  \caption{The nonplanar two-loop double box diagram with three off-shell legs.}
  \label{fig:offshell_nonplanar_dbox}
\end{figure}

Our first example is a 2-loop diagram shown in Fig.~\ref{fig:offshell_nonplanar_dbox}
with $5$ variables that we already used in \cite{Mokrov:2023vva}.
The legs $p_1$, $p_2$ and $p_3$ have mass $m$ while the leg $p_4$ is
massless. We perform IBP reduction of the three tensor integrals with
numerators $(k_1 + p_1)^2$, $(k_2 - p_4)^2$ and $(k_1 + p_1)^2 (k_2 - p_4)^2$, respectively.

We run a number of tests restricting the possible level of non-zero master integrals and thus modding the complexity of the reduction. We obtained the following results on a 4-core AMD Ryzen 7 3750H laptop:

\begin{table}[ht]
\caption{Reduction times (in seconds) for the two-loop nonplanar double box.}
\label{tab:npdbox}
\centering
\begin{tabular}{lcccc}
\hline
\textbf{Level of masters} & \textbf{7} & \textbf{6--7} & \textbf{5--7} & \textbf{4--7} \\
\hline
\texttt{Flint}                              & 22 & 896 &  13131 & RAM \\
\hline
\texttt{Flint + decomposed}                 & 14 & 726 &  11769 & RAM \\
\hline
\texttt{Flint + decomposed + factorized}    & 10 & 396 &  4809 & 16361 \\
\hline
\texttt{Symbolica}                          & 20 & 1151 & 13725 & RAM \\
\hline
\texttt{Symbolica + factorized}             & 23 & 890 &  4753 & 10643 \\
\hline
\end{tabular}
\end{table}

In this example symbolica in factorized mode becomes faster when a larger set of master integrals is used. However it one of out following examples the number of masters is even larger, and the results are opposite. RAM in the table means that 24Gb on the laptop was insufficient.

\subsubsection{Nonplanar double pentagon diagram}

Our second example is one of the $5$-variable examples we used in our {\tt FIRE} paper \cite{Smirnov:2025prc}.

\begin{figure}
  \centering
  \includegraphics[width=0.4\textwidth]{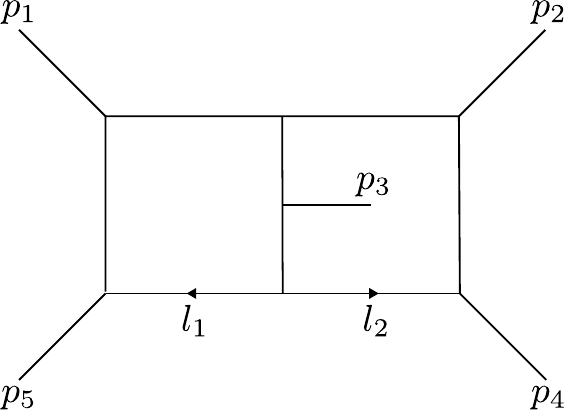}
  \caption{The nonplanar double pentagon diagram. The list of eight propagator denominators and three irreducible scalar products are $l_1^2$, $(l_1-p_5)^2$, $(l_1-p_5-p_1)^2$, $(l_2-p_4-p_2)^2$, $(l_2-p_4)^2$, $(l_2)^2$, $(l_1+l_2)^2$, $(l_1+l_2+p_3)^2$, $(l_2+p_5)^2$, $(l_1+p_4)^2$, and $(l_1+p_4+p_2)^2$.}
  \label{fig:nonplanarDoublePentagon}
\end{figure}

For the reason of a faster benchmarking we restricted the reduction to the top sector and ran
a number of tests reducing an integral with different powers of irreducible denominators.
The reduction was run on a cluster node using AMD EPYC 74F3 24-Core Processors with 16 cores reserved for the job.

In this test {\tt flint} with factorization of denominators showed a better performance than symbolica.

\begin{table}[ht]
\caption{Reduction times (in seconds) for the nonplanar double pentagon diagram.}
\label{tab:doublePentagon}
\centering
\begin{tabular}{lcccc}
\hline
\textbf{Numerators power} & \textbf{1} & \textbf{2} & \textbf{3} & \textbf{4} \\
\hline
\texttt{Flint}                              & 2412 & 4271 & 11313 & 30668 \\
\hline
\texttt{Flint + decomposed}                 & 1575 & 2782 & 7891 & 23321 \\
\hline
\texttt{Flint + decomposed + factorized}    & 900  & 1498 & 3915 & 12132 \\
\hline
\texttt{Symbolica}                          & 3390 & 5678 & 15066 & 39891 \\
\hline
\texttt{Symbolica + factorized}             & 1722 & 2645 & 7308 & 20079 \\
\hline
\end{tabular}
\end{table}

\subsubsection{Three-loop window}

\begin{figure}[h]
  \centering
  \includegraphics[width=0.44\textwidth]{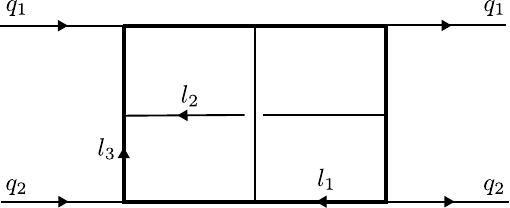}
  \caption{The three-loop forward-scattering diagram with massive internal lines, with $q_1^2=q_2^2=0$, $(q_1+q_2)^2 = s$. The list of 10 propagators are $l_3^2 - m^2$, $(l_2+l_3)^2 - m^2$, $l_1^2 - m^2$, $(l_3-q_2)^2 - m^2$, $(l_2+l_3+q_1)^2 - m^2$, $(l_1+q_2)^2-m^2$, $(l_1+l_2+q_1+q_2)^2-m^2$, $(l_1+l_2+q_2)^2 - m^2$, $l_2^2$, and $(l_1-l_3+q_2)^2$. The two ISPs are $(l_1+l_2+l_3)^2$ and $(l_1-q_1)^2$.}
  \label{fig:threeLoopWindow}
\end{figure}

We took the third example from our {\tt FIRE7} paper, section 5.3. --- a 3-loop
integral family with massive internal lines. The kinematics is exact
forward scattering with massless incoming momenta,
$q_1 + q_2 \rightarrow q_1 + q_2$. The diagram is shown in
Fig~\ref{fig:threeLoopWindow}. We reduce only sample integrals in
this benchmark, where the 4th propagator $(l_3-q_2)^2 - m^2$ is raised
to a 2nd power, and the total power of numerators varies.

We obtained the following results on the same 4-core AMD Ryzen 7 3750H laptop:

\begin{table}[ht]
\caption{Reduction times (in seconds) for the three-loop forward-scattering diagram.}
\label{tab:threeLoopWindow}
\centering
\begin{tabular}{lcccc}
\hline
\textbf{Level of numerators} & \textbf{0} & \textbf{1} & \textbf{2} & \textbf{3} \\
\hline
\texttt{Flint}                              & 1598 & 5133 & 6884 & RAM \\
\hline
\texttt{Flint + decomposed}                 & 1045 & 3298 & 4681 & RAM \\
\hline
\texttt{Flint + decomposed + factorized}    & 1030 & 3290 & 4352 & 16826 \\
\hline
\texttt{Symbolica}                          & 1131 & 3325 & 5016 & RAM \\
\hline
\texttt{Symbolica + factorized}             & 1002 & 3183 & 4422 & 15886 \\
\hline
\end{tabular}
\end{table}

Here we see that {\tt FLINT} works approximately at the same speed with {\tt Symbolica} with the factorized approach. The gain of the factorization is small, but the RAM gain is also efficient --- I did not have anough RAM for non-factorized run. It is also important that the decomposed output mode is essential compared with pure {\tt FLINT}.

\subsection{Internal code measurements}

Our goal was also to verify that out code is properly optimized, so we performed some internal performance measurements. For a big enough example we chose the two-loop massive nonplanar double box from section~\ref{subsec:npdbox} with master integrals of levels $5-7$. We ran a measurements with {\tt google perf}. Our goal was to measure the time spent in the {\tt fuel::base::evaluate} function called by {\tt FIRE} for expression simplification. The time usage based on collected stack data is the following:

\begin{itemize}
 \item 34.30 percent are calls to addition;
 \item 56.19 percent are calls to multiplication;
 \item 1.08 percent are calls to division (where factorization happens);
 \item The remaining part consists of string, memory operations and expression output.
\end{itemize}

We also tested this on examples of different size and clearly see that the division and factorization become negligible when the expression is large enough.

As for the large parts for addition and multiplication we demonstrated above how our algorithms perform them. The measurements on this example also show that we have small overhead costs.

The addition can be split as

\begin{itemize}
 \item 31.07 percent for calls to {\tt FLINT} multiplication;
 \item 45.52 percent for calls to {\tt FLINT} division (or gcd);
 \item 6.34 percent for calls to {\tt FLINT} addition;
\end{itemize}

The multiplication can be split as

\begin{itemize}
 \item 75.28 percent for calls to {\tt FLINT} multiplication;
 \item 18.99 percent for calls to {\tt FLINT} division (or gcd);
\end{itemize}

This measurement clearly shows that the overhead costs are quite small and that the implementation mostly relies on our algorithms (which we believe to be effective) and the addition, multiplication and division algorithms inside the {\tt FLINT} library which are well-tested for years. Thus we cannot expect the performance to be improved much more with the current approach.

\section{Conclusion}
\label{sec:conclusion}

In this work we proposed to exploit the factorized structure of denominators that naturally appears in integration-by-parts reductions.
We introduced a decomposed output format for polynomials and rational functions and extended it to support denominators stored as products of simple polynomials raised to integer powers, together with a corresponding serialization and parsing scheme.
On top of this representation we implemented arithmetic operations that use the factorized form explicitly, avoiding unnecessary polynomial multiplications and enabling systematic cancellation of common factors already at the level of the serialized data. The resulting functionality is integrated into {\tt FUEL} and can be used transparently from {\tt FIRE} for problems with many kinematic invariants.
Further work will focus on a more extensive performance study on realistic multi-scale IBP reduction.

\section{Acknowledgments}
The work of Alexander Smirnov was supported by the Ministry of Education and Science
    of the Russian Federation as part of the program of the Moscow Center for Fundamental
    and Applied Mathematics under Agreement No.\ 075-15-2025-345.

    We are grateful to Mao Zeng for assisting with examples and benchmarks for this article.

\clearpage

\bibliographystyle{unsrt}

\end{document}